\documentclass[
aps,
tightenlines,
nofootinbib,%
superscriptaddress,%
amsfonts,amsmath]{revtex4}

\usepackage{color}
\newcommand{\ba}{\begin{eqnarray}}
\newcommand{\ea}{\end{eqnarray}}
\newcommand{\nn}{\nonumber}
\newcommand{\be}{\begin{equation}}
\newcommand{\ee}{\end{equation}}

\usepackage{graphicx}

\begin{document}

\title{Cosmic strings\footnote{Review published in Scholarpedia \cite{Vachaspati:2015}, {\tt http://www.scholarpedia.org/article/Cosmic\char`_strings} }}

\author{Tanmay Vachaspati} \email{tvachasp@asu.edu}
\affiliation{Physics Department, Arizona State University, Tempe, Arizona 85287, USA.}

\author{Levon Pogosian} \email{levon@sfu.ca}
\affiliation{Department of Physics, Simon Fraser University, Burnaby, BC, V5A 1S6, Canada}

\author{Dani\`ele A.~Steer} \email{steer@apc.univ-paris7.fr}
\affiliation{AstroParticule \& Cosmologie,
UMR 7164-CNRS, Universit\'e Denis Diderot-Paris 7,
CEA, Observatoire de Paris,
10 rue Alice Domon et L\'eonie
Duquet, F-75205 Paris Cedex 13, France}
\affiliation{${\mathcal{G}}{\mathbb{R}}\varepsilon{\mathbb{C}}{\mathcal{O}}$,
Institut d'Astrophysique de Paris,
CNRS, UMR 7095, et Sorbonne Universit\'es,
UPMC Univ Paris 6,
98bis boulevard Arago, F-75014 Paris, France}
\affiliation{LAPTH, Universit\'e Savoie Mont Blanc, CNRS,
B.P.110, F-74941 Annecy-le-Vieux Cedex, France}

\begin{abstract}
This article, written for Scolarpedia \cite{Vachaspati:2015}, provides a brief introduction into the subject of cosmic strings, together with a review of their main properties, cosmological evolution and observational signatures.

\end{abstract}

\date{\today}


\maketitle

\tableofcontents

\section{Introduction}
``Strings'' are solutions of certain field theories, whose energy is concentrated along an infinite line.  Strings exist in many field theories motivated by particle physics, and this suggests that they may exist in the universe --- hence the name ``cosmic strings''. String solutions are also present in condensed matter systems where they are called ``vortices''.
In cosmological applications, strings are generally curved, dynamical, and may form closed loops.  The energy of a string remains concentrated along a time-dependent curve for a duration that is very long compared to the dynamical time of the string.

\section{Role of Topology}
\label{Sec1}

The topological properties of a field theory may be used to motivate the existence of string solutions. If a field theory has certain symmetries and symmetry breaking patterns, the vacuum state (the state of lowest energy) may not be unique. The collection of possible vacua form a manifold, $M$, which may have ``holes" i.e.~there may be closed paths on $M$ that cannot be continuously shrunk to a point. In this case, the field theory has topology that is suitable for the existence of string solutions. In mathematical terms, the topology relevant for strings is described by the first homotopy group of the vacuum manifold, $\pi_1 (M)$. The relevance of topology is best understood with an example.

\begin{figure}[h]
\includegraphics[angle=0,width=10cm]{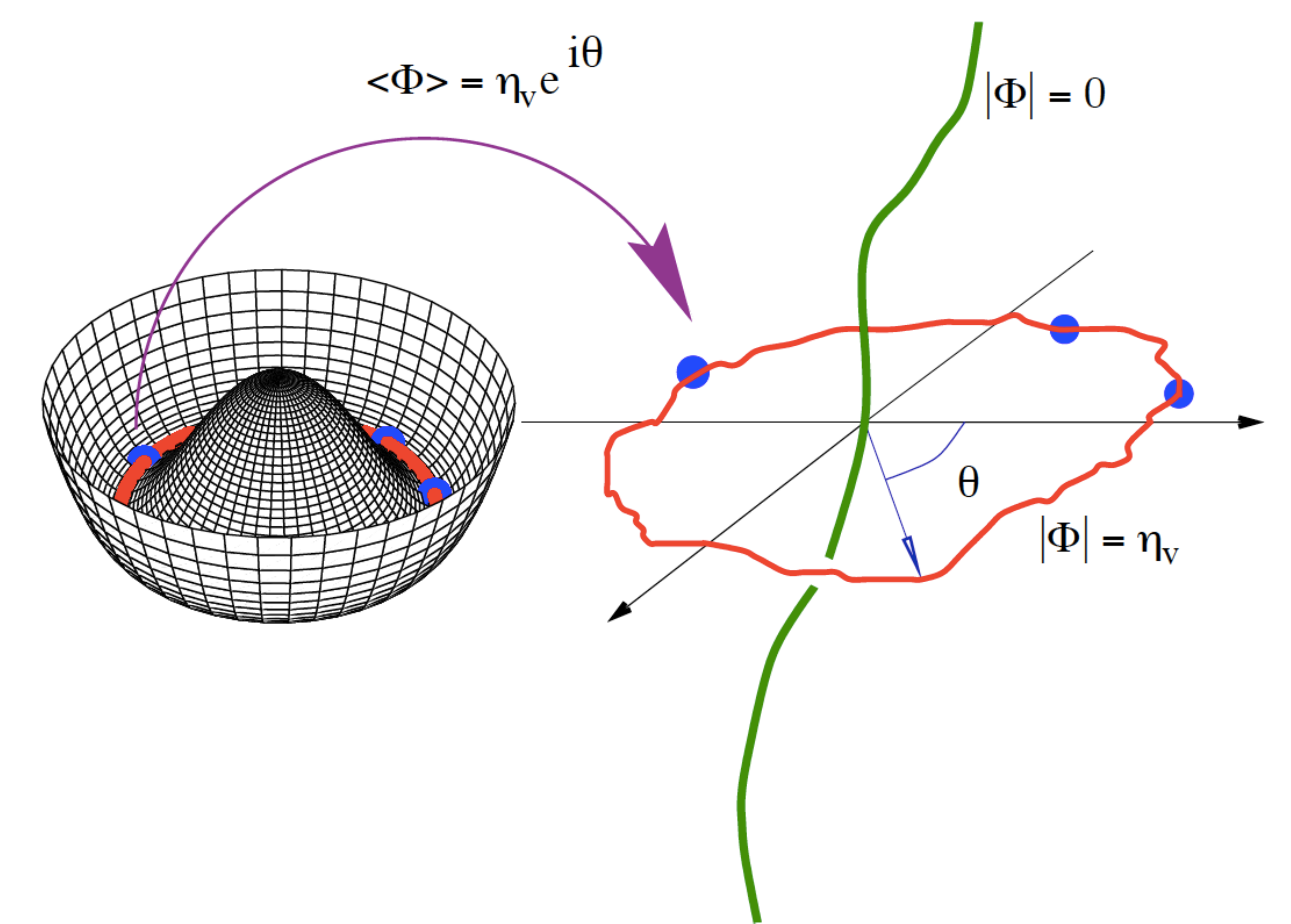}
\caption{String formation in the "Mexican-hat" potential $V(|\Phi|)$. The potential is shown on the left-hand-side, with its circular manifold (red) on which 3 points (blue) have been chosen at random.  The right-hand-side shows (in red) a closed path in ''physical space'', along which $|\Phi|=\eta_v$; the blue points in physical space are the points that map on to the blue points shown on the vacuum manifold. On going around the path in physical space, the field wraps once around the vacuum manifold .  By continuity of the field, $\Phi$ must vanish somewhere within the circle in physical space. This is the center of the string drawn in green.  Fig.~ from \cite{Ringeval:2010caca}.}
\label{fig1}
\end{figure}

Consider a complex scalar field, ${\Phi}$, in three spatial dimensions, with potential energy function, 
\be
V(|\Phi|)=(|\Phi|^2 -\eta_v^2)^2
\nn
\ee
 The minimum energy configuration has $|{\Phi}|=\eta_v$ but the phase of $\Phi$ is undetermined and labels the points on the vacuum manifold which is a circle. A closed path that wraps around the circle cannot be continuously contracted to a point and hence there can be strings in this field theory. If, as one goes around a closed path in physical space, one also wraps around around the circle on the vacuum manifold $n=\pm 1, \pm 2,\ldots$ times, then there will be possibly $n$ strings going through the closed path in physical space. (See Fig.~\ref{fig1}.)  Notice that at the center of the string $|\Phi|=0$ and hence the energy density is non-zero at the string core.  

\noindent {\bf Caution:}  Non-trivial topology of a field configuration does not necessarily imply the existence of a {\it static} solution. For example, if two $n=1$ strings repel at all separations, then an $n=2$ configuration will split into two $n=1$ strings that move apart. In this case, there is non-trivial topology since $n=2$, but no static solution exists.

\section{Solution and properties}
\label{Sec2}
\subsection{Straight global string}
For a straight, static string, it is sufficient to look for a solution of the equations of motion in two spatial dimensions, and then use translation invariance to extend the solution to three dimensions. For example, if the solution in two dimensions is $\Phi_0(x,y)$, then the solution in three dimensions is $\Phi(x,y,z)=\Phi_0(x,y)$.

The simplest theory which gives rise to string solutions is described by the Lagrangian
$$
L = \left |\partial_\mu \Phi \right |^2 - \frac{\lambda}{4}(|\Phi|^2-\eta^2)^2
$$
where $\lambda$ is a dimensionless coupling constant, $\eta$ is the vacuum expectation value of the field $\Phi$, and the metric has signature $(+,-,-,-)$. We will also use natural units throughout so that $\hbar=c=1$. This Lagrangian is invariant under a global $U(1)$ symmetry, $\Phi \rightarrow \Phi e^{i\Lambda}$ (for any constant $\Lambda$), and the corresponding equations of motion are
$$
\partial_\mu \partial^\mu \Phi =-\lambda (|\Phi|^2 - \eta^2) \Phi.
$$

The static string solution in this model is
$$
\Phi(x,y)=\eta \, f(m \rho) e^{in \theta}
$$
where $(\rho,\theta)$ are polar coordinates on the $xy$-plane, $m^2 = \lambda \eta^2$, and $n$ is the (integer) winding number of the string. On substituting into the equations of motion, the function $f(\tilde{\rho})$, ${\tilde \rho}\equiv m \rho$, has the features
\ba
f(\tilde{\rho}) = f_0 \tilde{\rho}^{|n|} (1+ {\cal{O}}(\tilde{\rho}^3)),&&  \ \qquad \tilde{\rho} \ll 1 
\nn
\\ 
f(\tilde{\rho})=1 - {\cal O}\left (\frac{1}{\tilde{\rho}^2}\right ),&& \ \qquad \tilde{\rho}  \gg 1.
\nn
\ea
The energy density ${\cal E} = |\vec{\nabla} \Phi|^2 + V(\Phi)$ is peaked within $\rho \sim m$, and falls off as $1/\rho^2$ at large distances. The total energy per unit length of the string diverges weakly (logarithmically). In a physical setting when there are lots of strings or in a condensed matter sample of finite volume, the divergence gets cut off. This string solution is known as a ``global" string because there are no gauge fields in the model.

\subsection{The Nielsen-Olesen string}
The model can be extended to include gauge fields 
$$
L =\left |D_\mu \Phi \right |^2 - \frac{1}{4} F_{\mu\nu}F^{\mu\nu} - \frac{\lambda}{4}(|\Phi|^2-\eta)^2
$$
where $D_\mu = \partial_\mu - ie A_\mu$ is the gauge covariant derivative, and $F_{\mu \nu} = \partial_\mu A_\nu - \partial_\nu A_\mu$. This ``Abelian Higgs model" was considered by Nielsen and Olesen in their discovery paper on string solutions in relativistic field theories 
\cite{Nielsen:1973cs}. The Lagrangian is now invariant under a local $U(1)$ symmetry in which $\Phi \rightarrow \Phi e^{i\Lambda(x)}$ and $A_\mu \rightarrow \partial_\mu \Lambda/e$.  In the static string configuration, the asymptotic properties of the scalar field differs from the global case. In particular as $\rho \rightarrow \infty$, the scalar and gauge fields both contribute to the energy density and make it fall off exponentially fast, and the energy per unit length $\mu$ of the string is finite. 

The string also contains a flux of magnetic field that is quantized as can be seen by noting that $D_\mu\Phi \to 0$ outside the string,
$$
{\rm Magnetic\ flux} = \oint dx^\mu A_\mu = \frac{2\pi n}{e}
$$

\subsubsection{Type I and Type II strings}

The properties of strings in the local $U(1)$ model depend on the ratio of coupling constants $\beta = \lambda/2e^2$.  In the limit $\beta=1$ the equations of motion simplify and  an important method to find string solutions,  also often used in supersymmetric field theories, was developed by Bogomol'nyi \cite{Bogomolny:1975de}.  Here the energy per unit length $\mu_n$ of a string with winding number $n$ is exactly equal to that of $n$ strings each with winding number $1$. In this case $\mu_1 = \pi \eta^2$
and $\mu_n = n \mu_1$.  

\begin{figure}[h]
\includegraphics[angle=0,width=10cm]{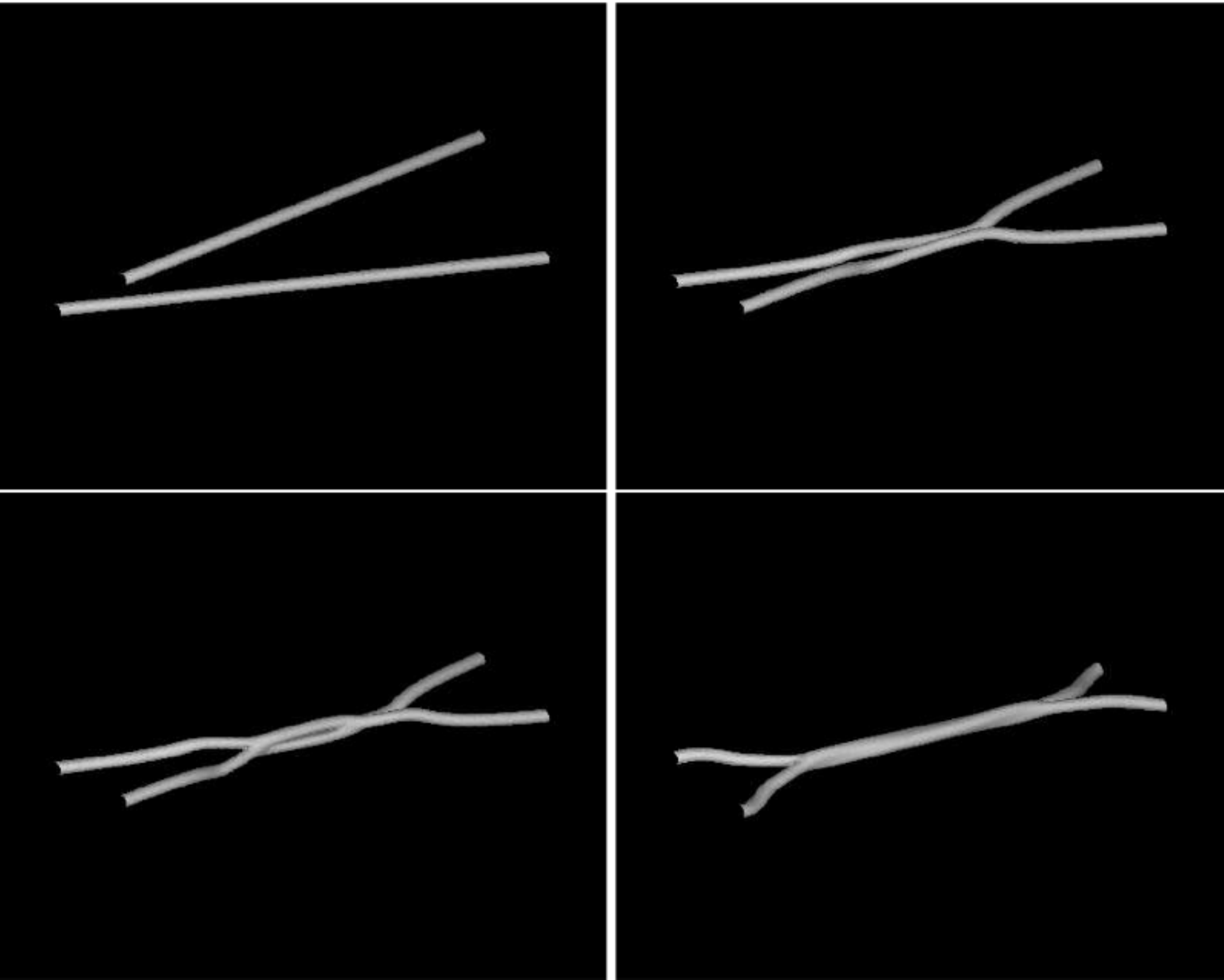}
\caption{Snapshots (from left to right and top to bottom) 
showing constant energy 
density isosurfaces in a simulation of two strings with $n=1$ in the type I regime ($\beta=0.125$) colliding to form an $n=2$ string . From \cite{Salmi:2007ah}.}
\label{fig2}
\end{figure}

In the limit $\lambda < 1$, often called the type I regime in analogy with superconductors, $\mu_n < n \mu_1$. In particular, two $n=1$ strings can merge to form 
an $n=2$ string \cite{Jacobs:1978ch}, see Fig.~ \ref{fig2}.  

\begin{figure}[h]
\includegraphics[angle=0,width=10cm]{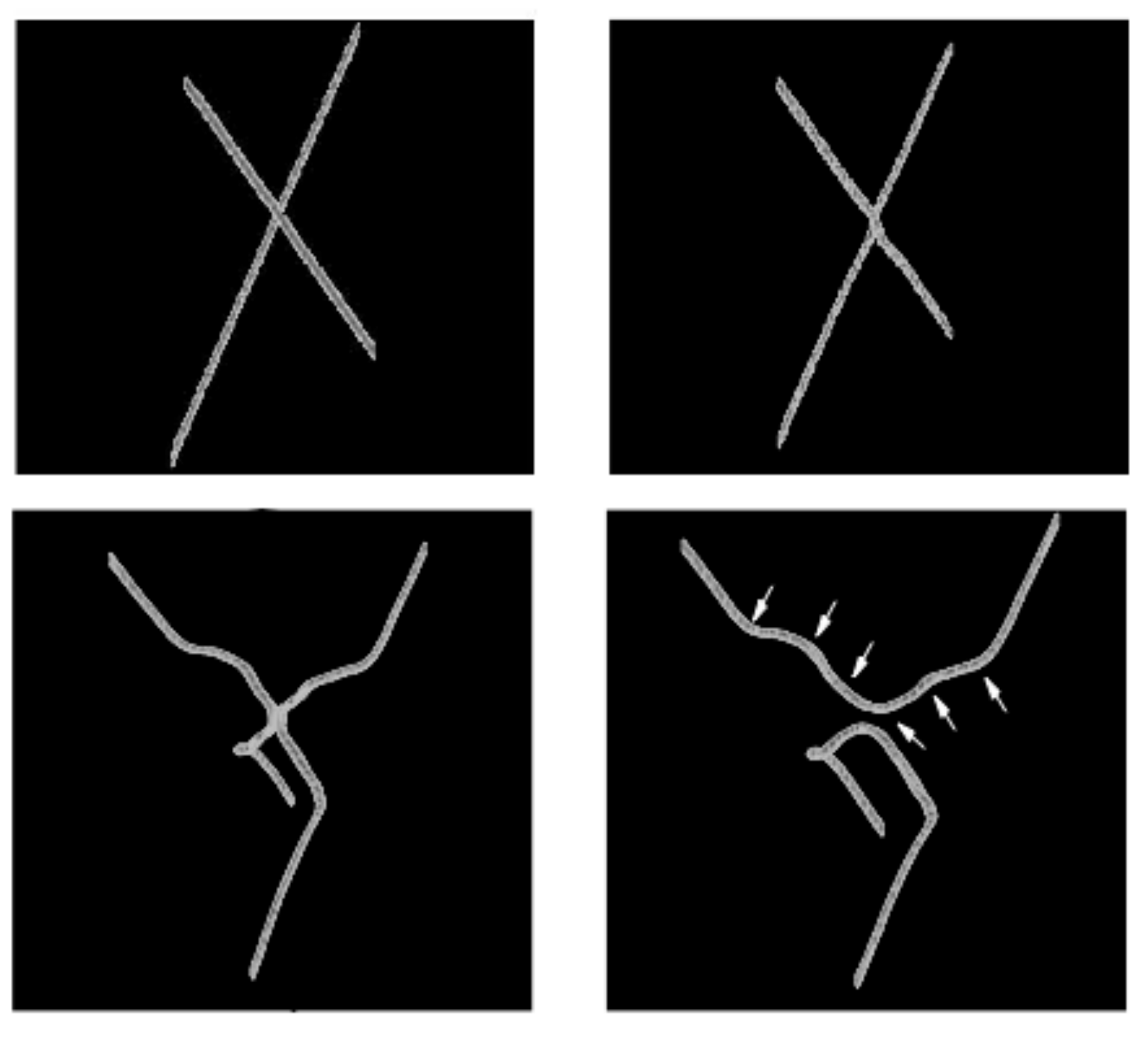}
\caption{Snapshots (from left to right and top to bottom) 
showing constant energy 
density isosurfaces in a simulation of two strings with $n=1$ in the type II regime ($\beta=32$) colliding and intercommuting with the formation of kinks (as indicated by arrows). From 
\cite{Verbiest:2011kv}.}
\label{fig3}
\end{figure}

In the type II regime, $\lambda > 1$ and only the winding 1 strings are stable.  Usually when discussing the cosmological properties of cosmic strings, the strings being considered are those of the local $U(1)$ model in the type II regime. These are referred to as ``gauge strings" or ``local strings".  When two type II strings collide, for essentially all angles and collision velocities, they ``intercommute": that is, they exchange partners (Fig.~\ref{fig3}).  Thus gauge strings have an intercommutation probability $P= 1$  \cite{Shellard:1987bv,Matzner1988}, except at very high incoming velocities \cite{Verbiest:2011kv}.

\subsection{Other types of strings}
If there are fermions in the model that couple to the scalar field that winds around the string, ``fermion zero modes" may exist \cite{Jackiw:1981ee}. These are solutions of the Dirac equation that are localized on the string and have zero energy. If the fermions also carry electromagnetic charge, the cosmic strings can carry electric currents, leading to interesting astrophysical signatures in the cosmological context. In some models, charged scalar fields can also be localized on the string. Current-carrying strings are also known as ``superconducting strings" \cite{Witten:1984eb}.

Many other types of strings (e.g. semi-local strings, Alice strings etc) can form depending on the topology and coupling to other fields (see \cite{Achucarro:1999it,Copeland:2011dx,Hindmarsh:1994re,Ringeval:2010ca,ViSh}).

In summary, the basic structure of a string is a scalar field that winds around the location of the string, where there is a concentration of energy density. Gauge fields that interact with the scalar field provide the string with a quantized magnetic flux. Fermion zero modes can be localized on the string and be responsible for currents that run along the string.

\section{Bulk Properties}

If the mass of the scalar field that winds around the string is $m$ and all dimensionless coupling constants are O(1), the width of a local string is $\approx m^{-1}$. In most cosmological applications, the width of the string is very small compared to the other length scales in the problem, and the thin string limit is commonly adopted.  Then the string is simply modeled as a line with mass per unit length $\mu \approx m^2$. If the string is not superconducting, its tension $T$, {\it i.e.~}the longitudinal component of the string energy-momentum tensor, is also $\mu$. In the zero-width approximation, the strings are referred to as ``Nambu-Goto" strings as their dynamics is obtained by solving the Nambu-Goto action which minimises the area swept out by the worldsheet of the string. Numerically, $\mu \approx 10^{22}~{\rm gms}/{\rm cm}$  and $G\mu \approx 10^{-6}$ when $m \approx 10^{16}~{\rm GeV}$. 

An important feature of Nambu-Goto strings is that they contain ``kinks" and ``cusps". A kink is a point at which the tangent vector of the string changes discontinuously, and kinks are formed when strings intercommute (Fig.~ 3). Kinks travel along the string at the speed of light. At a cusp, the string instantaneously travels at the speed of light.  Kinks and cusps give rise to important observational signatures of strings (see below).

Superconducting strings can carry an electric current $j$ which can be timelike, spacelike or lightlike, and leads to an equation of state of the string $T=T(\mu)$.  In general, the maximum current allowed on the string is $O(m^2)$ though, depending on the detailed particle interactions, it can be substantially weaker \cite{Barr:1987ij,Peter:1992dw}.  The effective action for superconducting strings is no longer the Nambu-Goto action.

The metric around a static infinitely straight Nambu-Goto string lying along the $z$-axis can be obtained by solving the Einstein equations.  It is "conical" on the plane transverse to the string, and the line element is \cite{Vilenkin:1981zs}
$$
ds^2 = dt^2-dz^2 - d\rho^2 - \rho^2 d\theta^2\ , \ \ 0\le \theta < 2\pi (1-4G\mu )
$$
where $G$ is Newton's gravitational constant. As is apparent, the metric is locally flat and the only non-trivial feature is that the angular coordinate $\theta$ lies in an interval that is less than $2\pi$. 
This particular form of the metric is central to many of the observational signatures of cosmic strings described below.

In physical applications, a whole network of strings is formed when the symmetry is broken, and individual strings can be infinitely long or in the shape of closed loops, and the network evolves in time. A curved string is a dissipative solution of the equations of motion. Loops will eventually decay into various forms of radiation including the scalar and gauge fields of which the string is formed, and gravitational waves through the coupling of the string to gravity; infinite curved strings will tend to straighten out. The dissipation time-scale is generally very long compared to the dynamical time of loops for long loops, so the string picture is useful. For example, a loop will oscillate $\sim (100 G\mu)^{-1}$ times, (that is $\sim 10^5$ times for the string tension of $G\mu\sim 10^{-7}$), before losing an $O(1)$ fraction of its energy to gravitational waves.

In certain field theories, strings networks can also have junctions --- namely points at which three strings meet. In particular this occurs in the Abelian-Higgs model when $\beta < 1$ as is shown in Fig.~\ref{fig2}.  Junctions also occur in more complicated models in which non-abelian symmetries are broken.  Cosmic superstring networks, predicted in fundamental superstring theories, also have junctions. There they are located at the meeting point between
fundamental F-strings, Dirichlet D-strings and a bound states of these two.  Cosmic superstrings are  known as $(p,q)$-strings, an abbreviation for $p$~F-strings 
and $q$~D-strings \cite{Copeland:2003bj,Dvali:2003zj}.  The effective action for strings with junctions is a set of three coupled Nambu-Goto actions \cite{Copeland:2006eh}, but it should be noted that the intercommutation probability of superstrings $P<1$. 

Cosmic strings also interact with the ambient cosmological medium. Particles scatter off the string with differential cross-section per unit length \cite{Everett:1981nj}
$$
\frac{d\sigma}{d\theta} = \frac{\pi}{2k (\log(k \delta))^2},
$$ 
where $k$ is the momentum of the particle transverse to the string and $\delta$ is the string width.
The cross-section is larger if there are particles that scatter by the ``Aharonov-Bohm" interaction \cite{Rohm,ABef,Alford:1988sj}, 
$$
\frac{d\sigma}{d\theta} = \frac{\sin^2(\pi\nu)}{2\pi k \sin^2(\theta/2)}
$$
where $2\pi\nu$ is the Aharonov-Bohm phase. Note that the scattering cross-sections only depend on the momentum of the incoming particle, and are insensitive to the mass scale of the string. The interaction of strings with ambient particles plays an important role in the early stages after a string network forms as it over-damps the string dynamics. However, as the universe expands, the density of ambient matter falls and particle interactions cease to be an important factor.

\section{Cosmology}
\subsection{Formation}
Based on our current understanding of particle physics, the vacuum structure may have topology that is suitable for the existence of string solutions. The mathematical existence of string solutions in a field theory, however, does not imply that they will be realized in a physical setting and additional arguments are needed to make the case that strings can be present in the universe \cite{Kibble:1976sj}. Essentially, during spontaneous symmetry breaking, different vacua are chosen in different spatial domains, and the non-trivial topology of the vacuum manifold then inevitably implies the presence of strings in cosmology. At formation, a large fraction  of the string network (roughly 80\% in the simplest formation models) is in infinite strings and the rest is in loops with a scale-invariant distribution \cite{Vachaspati:1984dz}.

Subsequently, the network relaxes under several forces that include the string tension, frictional forces due to ambient matter, cosmic expansion, and the process of intercommuting.  In particular when a loop or an infinite string intercommutes with itself, it chops off a loop. This means that at any given time the network will contain many loops: those formed at time $t$ as well as at all previous times.  In addition, a Nambu-Goto loop evolves periodically in time and hence loses energy to gravitational and other forms of radiation. As a result a loop of initial length $\ell_0$ formed at time $t_0$ has a length $\ell (t)$ at time $t$ given by
$$
\ell (t) = \ell_0-\Gamma G\mu (t-t_0)
$$
where $\Gamma$ is a constant which, in the case of gravitational radiation, is of order 100 \cite{Vachaspati:1984gt} (the precise value depends on the shape of the loop).  A typical loop will have a number of kinks and cusps, and the spectrum of high-frequency gravitational radiation emitted from a string depends on these features.

\subsection{Evolution}

\begin{figure}[htb]
\includegraphics[angle=0,width=14cm]{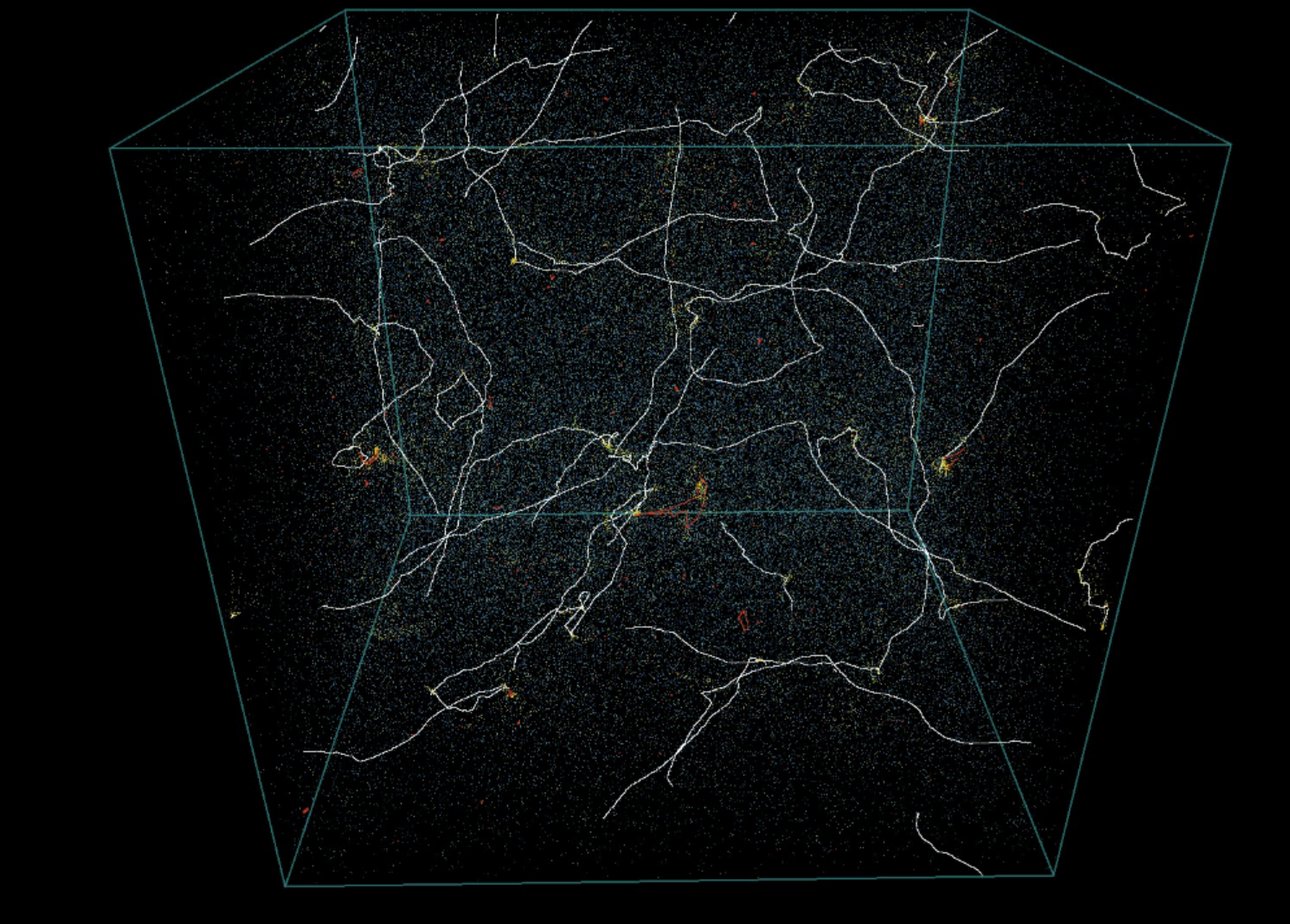}
\caption{Nambu-Goto strings in the matter era. Infinite strings (defined as having a size bigger than the conformal horizon, which is the box size in this figure) are colored in white. The red loops are those whose average density has reached scaling. The remaining loops are not (yet) in scaling, and they are color-coded according to their age: freshly formed loops are yellow and older loops are bluer. The bluest ones are therefore those coming from the initial conditions and the greenish ones have been formed in between. Figure given with kind permission by Christophe Ringeval.}
\label{fig4}
\end{figure}

The evolution of the network from its formation until today is an extremely complex problem involving very disparate length scales. Several groups have tackled this problem by using
numerical simulations of Nambu-Goto (zero thickness limit) strings, starting with the pioneering work in \cite{Albrecht:1989mk,Bennett:1989yp,Allen:1990tv}. Other groups have performed field theory simulations in which the strings have structure.
And yet others have built analytical models to describe the evolution of the network.
These analyses show that the network reaches a self-similar attractor solution on large scales
in which all the properties and length scales describing the network scale with time.  In this 'scaling solution', the physical number density of loops $n(\ell,t) d\ell$  with length between $\ell$ and $\ell+d\ell$ at cosmic time $t$ is characterized by different power-law behaviors. 
For example, recent simulations by \cite{Blanco-Pillado:2013qja}, give
$$
{n(\ell,t)} =  \frac{0.18}{t^{3/2} (\ell + \Gamma G\mu t) ^{5/2}},\ \ \ \ell < 0.1 t \ \ \ ({\rm radiation~ era})
$$
$$
{n(\ell,t)} =  \frac{0.27-0.45(\ell/t)^{0.31}}{t^{2} (\ell + \Gamma G\mu t)^{2}},\ \ \ \ell < 0.18 t \ \ \ ({\rm matter~ era}).
$$
Nambu-Goto simulations by other groups (see for instance \cite{Lorenz:2010sm}) find similar loop distributions though there is disagreement about very small loops that, however, carry only a very small fraction of the total energy in the network. In Abelian-Higgs simulations, many fewer loops are seen and the string network energy is mostly dissipated directly into particle radiation \cite{Vincent:1997cx}.

Notice that the above are the pure scaling distributions, meaning that the dimensionless number density of loops $t^4 n(\ell,t)$ only depends on the dimensionless ratio $\ell/t$ for all times. At formation though, the loops are not in the scaling distribution: they relax towards scaling after a time which can be estimated.
Numerical simulations, however, observe a population of non-scaling loops. Some of these are a remnant of the initial loop distribution formed at the phase transition, and others are small loops freshly formed from small scale structure on long strings (see Fig.~4). Similarly, on entering the matter era, the radiation era scaling distribution relaxes to the matter era scaling distribution. The timescale for this process depends on the length of the loop, and is longer for shorter loops.  

In addition to loops the network contains many infinite strings with typical separation $\sim 0.15 d_h$ where $d_h$ is the horizon distance \cite{BlancoPillado:2011dq}. This corresponds to $\sim 40$ infinite strings in any horizon volume. A typical distribution of strings is show in Fig.~\ref{fig4}.

\section{How we can look for them}
The presence of strings in the universe can be deduced from their gravitational effects and other non-gravitational signatures if they happen to couple to other forces. For example, cusps on cosmic string loops emit bursts of gravitational waves \cite{Damour:2000wa}. Moving strings produce wakes in matter and line discontinuities in the cosmic microwave background (CMB). They also induce characteristic patterns of lensed images of background light sources. Superconducting strings, in addition to the above effects, emit electromagnetic radiation that can potentially be detected as radio bursts. At present, the strongest bounds on the string tension come from constraints on the stochastic gravitational wave background from pulsar timing measurements and the LIGO interferometer. However, these bounds are sensitive to the details of the string network evolution. On the other hand, bounds from CMB are weaker but also less model-dependent. Different types of cosmic string signatures and their current status are reviewed below.

\subsection{Gravitational signatures}
\subsubsection{Cosmic Microwave Background}

\begin{figure}[htb]
\includegraphics[angle=0,width=14cm]{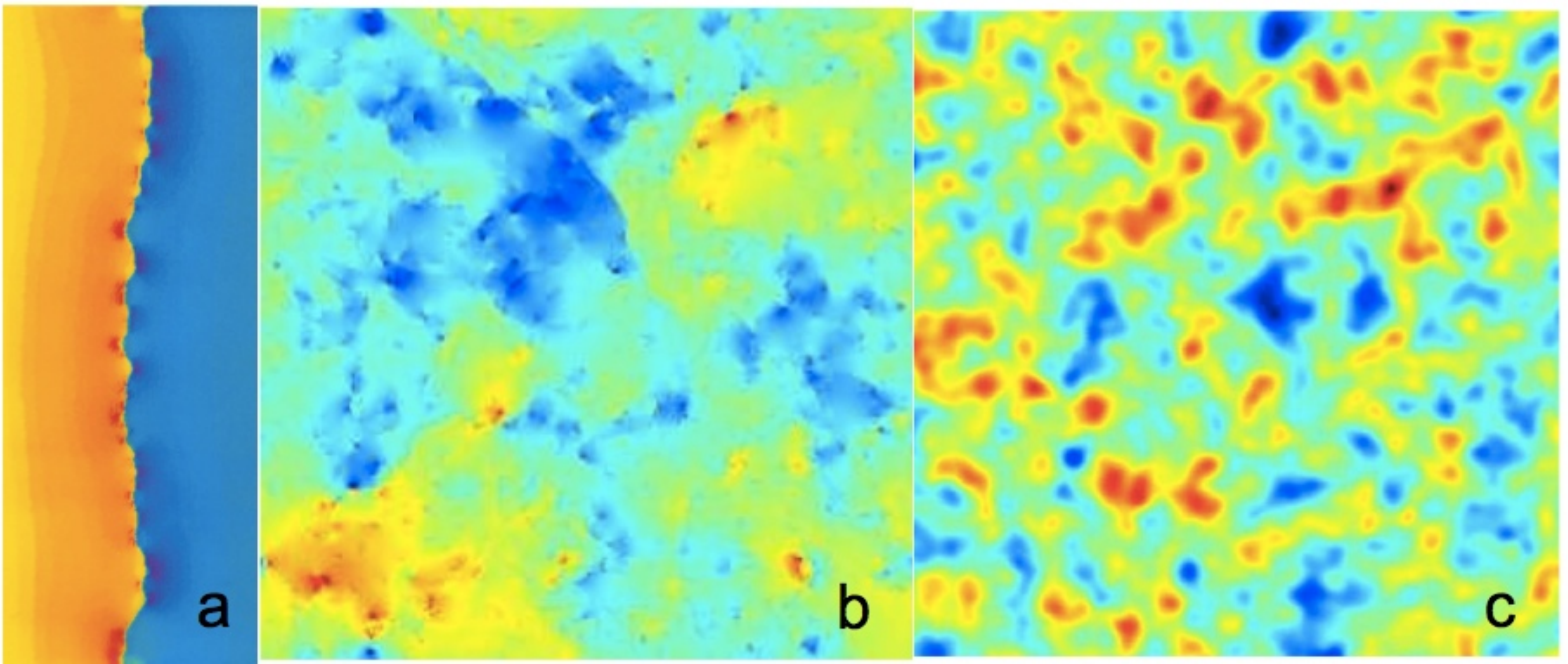}
\caption{CMB anisotropy sourced by strings: a) a line-discontinuity in CMB temperature caused by a single string on a uniform background (image provided by Proty Wu and Paul Shellard, (J.H.P.Wu PhD thesis, U. of Cambridge, 2000)); b) anisotropy caused by a network of strings via the Kaiser-Stebbins-Gott effect alone (image from \cite{Fraisse:2007nu}); c) anisotropy caused by a network of strings with full recombination physics taken into account. (Image from \cite{Landriau:2010cb}.)}
\label{fig5}
\end{figure}

Cosmic string networks persist throughout the history of the universe and actively source metric perturbations at all times. Prior to cosmic recombination, density and velocity perturbations of baryon-photon fluid are produced in the wakes of moving cosmic strings, which then remain imprinted on the surface of last scattering.  After recombination, strings crossing our line of sight generate line-like discontinuities in the CMB temperature, which is the so-called Kaiser-Stebbins-Gott (KSG) effect \cite{Kaiser:1984iv,Gott:1984ef,Bouchet:1988hh}. Both, wakes and the KSG effect, are induced by the deficit angle in the metric around a string. In addition, matter particles experience gravitational attraction to the string if it is not perfectly straight. The search for cosmic string signatures in the CMB can be broadly divided into attempts to {\it directly} detect line discontinuities in the temperature or polarization patterns, and {\it statistical methods} based on calculations of various correlation functions.

{\it Direct Searches.} The spacetime around a straight cosmic string is locally flat, but globally conical, with a deficit angle determined by the string tension. Thus, a string passing across our line of sight would produce a discrete step in the CMB temperature proportional to $G \mu \left|{\vec v} \times {\hat n}\right | $, where 
${\vec v}$ is the velocity of the string and ${\hat n}$ is the direction of the line of sight (see Fig.~ 5). Several groups have tried searching for such line-like features in the existing CMB maps and to forecast the prospects for future observations.
The tightest existing bound on the string tension based on direct signatures is  $G\mu < 3.7 \times 10^{-6}$ at the 95\% confidence level \cite{Jeong:2004ut} and is based on the assumption that the number density of strings is approximately known. The most optimistic forecast, based on Canny algorithm, claims that direct searches with future CMB experiments can achieve bounds of $G\mu < 3 \times 10^{-8}$ \cite{Danos:2008fq}. 
Detectable sharp edges can be present not only in CMB temperature maps, but also in {\it polarization} maps. The primary limitation in these types of studies comes not so much from the instrumental noise and angular resolution of the experiment, but from the fact that CMB is dominated by the Gaussian fluctuations on scales comparable to the size of the horizon at decoupling. Also, the above mentioned forecasts assume idealized line discontinuities produced by straight string segments. Actual strings are not straight, and contain both infinite strings and string loops. It remains to be seen how well these methods perform under more realistic assumptions.

\begin{figure}[htb]
\includegraphics[angle=0,width=10cm]{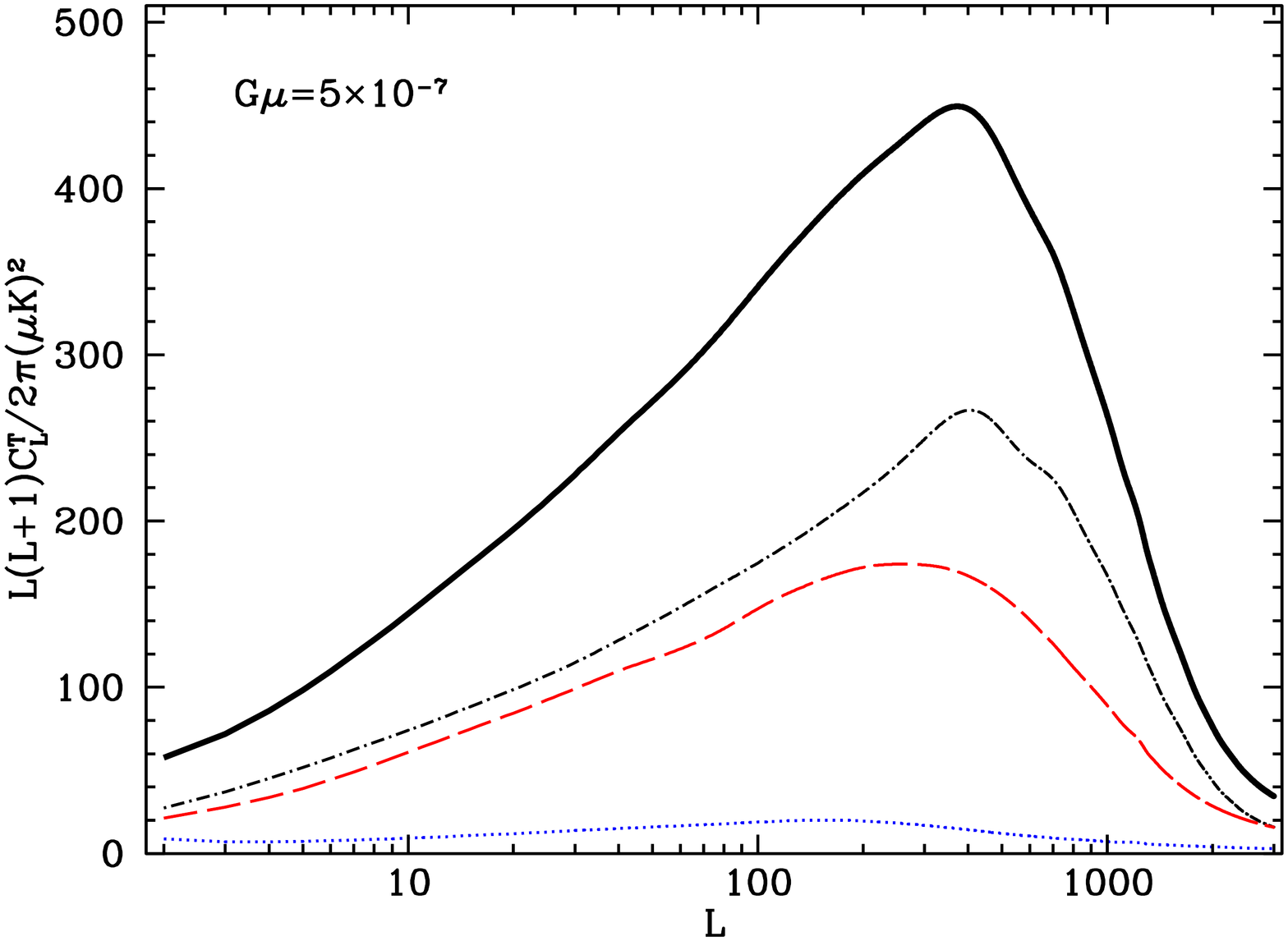}
\includegraphics[angle=0,width=10cm]{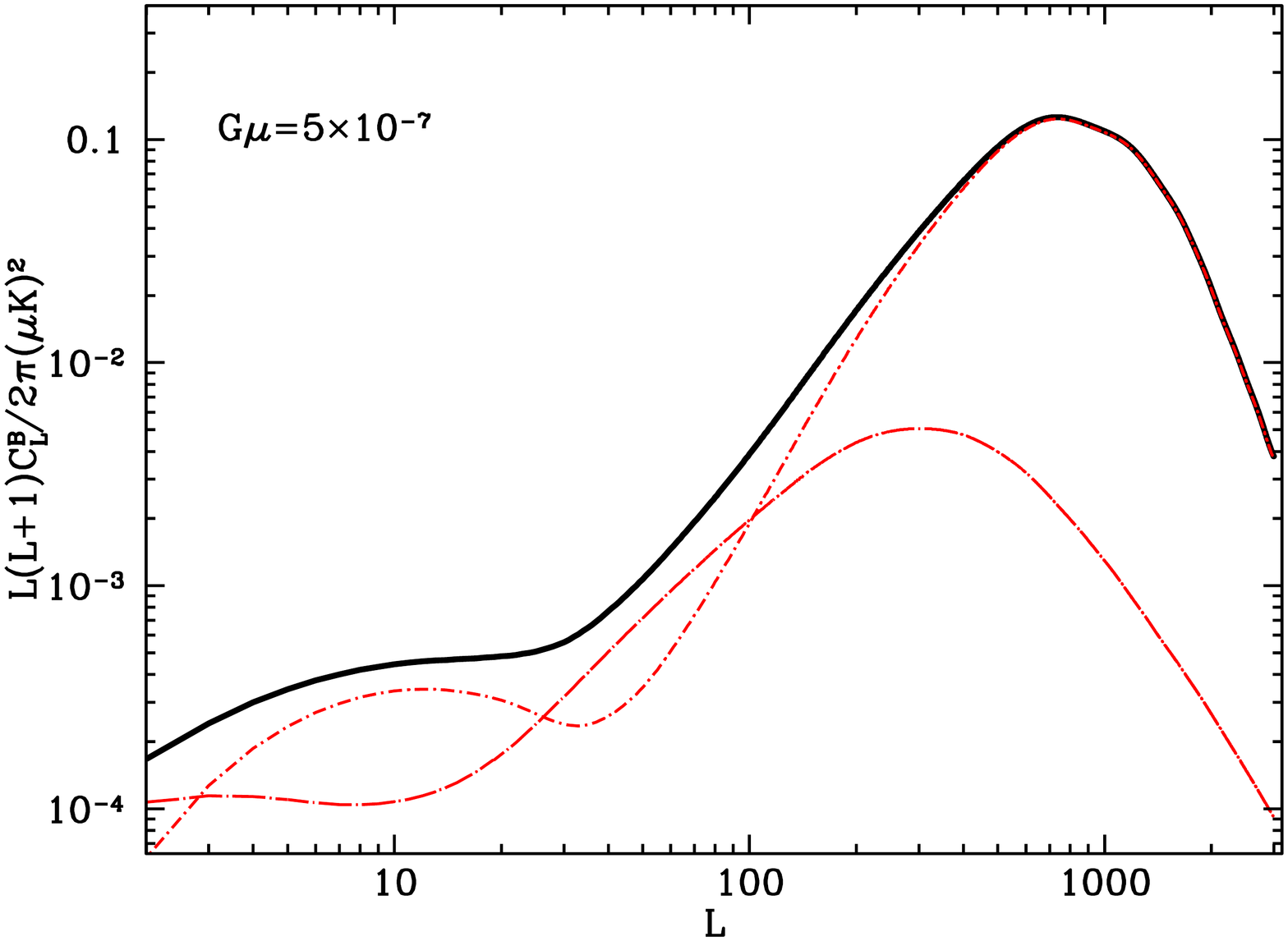}
\caption{The CMB temperature (top) and B-mode polarization (bottom) spectra sourced by strings. Contributions from scalar (black dash-dot), vector (red dash) and tensor modes (blue dot) and their sum (solid black) are shown separately.}
\label{fig6}
\end{figure}

{\it Statistical Methods.} The most reliable bounds on cosmic strings are derived from the angular power spectrum of CMB temperature anisotropies measured by the WMAP and Planck satellites. Calculating the spectrum of CMB anisotropies sourced by strings requires tracking their evolution from just before recombination until today over a large range of scales. Because obtaining an exact solution is quite challenging numerically, several approximate and semi-analytical methods have been used to evaluate the string CMB spectra. There is broad agreement between results from different approaches on the general shape of CMB spectra sourced by ``local'' cosmic strings, with the allowed fraction of the string contribution to the total CMB temperature anisotropy currently limited to under 3\% \cite{Ade:2013xla}. For ``conventional'' strings, i.e. those with order unity inter-commutation probability, this implies $G\mu \lesssim few \times 10^{-7}$. 

While most approaches adopt the Nambu-Goto approximation, \cite{ Bevis:2006mj}, evolved the cosmic string field configurations in the Abelian-Higgs model. To make the calculation of CMB spectra numerically feasible, the fields were separately evolved over limited time ranges in the radiation and matter eras, and an interpolation scheme was used to connect the two scaling regimes. Perfect scaling was then assumed to extend their range to later times. To circumvent the problem of resolving the fixed width core of strings in an expanding background, the core size was allowed to grow with the expansion in a prescribed fashion. This approach is designed to directly calculate the spectra and cannot predict CMB maps.

Another method predicting the spectra but not the map, is the so-called Unconnected Segment Model implemented in the publicly available code CMBACT \cite{Pogosian:1999np,CMBACT}. In the unconnected segment model, the string network is represented as a collection of uncorrelated straight string segments.

\begin{figure}[htb]
\includegraphics[angle=0,width=12cm]{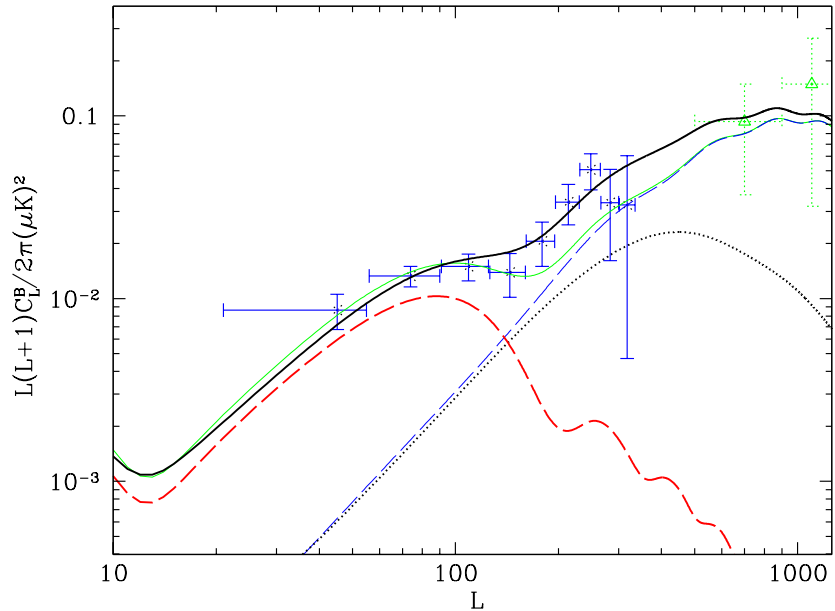}
\caption{B-mode spectra from cosmic strings (black dot), weak lensing (blue long dash), inflationary gravity waves with r=0.15 (red dash) and their sum (black solid) along with the BICEP2 \cite{Ade:2014xna} and POLARBEAR \cite{Ade:2014afa} data points. The faint solid line is the BICEP2 best fit model with r=0.2.}
\label{fig7}
\end{figure}

Although strings cannot be the main source of the CMB temperature anisotropy, they can generate observable B-mode polarization. The B-mode from strings is primarily generated by vector modes, with a spectrum that is different from the one generically produced from tensor modes arising in inflationary scenarios. Future CMB polarization experiments should be able to reveal the presence of cosmic strings through their B-mode signature even if strings contribute as little as $0.1\%$ to the CMB temperature anisotropy. Constraints comparable to those previously obtained from CMB temperature spectra were obtained with the POLARBEAR \cite{Ade:2014afa} and BICEP2 \cite{Ade:2014xna}  B-mode spectra in \cite{Moss:2014cra,Lizarraga:2014eaa,Lizarraga:2014xza,Lazanu:2014eya}. Given the uncertainty regarding the potentially significant unresolved foreground contributions to the BICEP2 signal, some of these results and conclusions may have to be revisited in the future.

 Typical CMB temperature and polarization spectra sourced by local cosmic strings are shown in Fig.~s \ref{fig6} and \ref{fig7}. The temperature spectrum features a broad peak at multipole moment $L \sim 300-500$ and no acoustic peaks. The B-mode spectrum has a peak at $L \sim 600$ and lower peak at $L \sim 10$ due to polarization generated at the epoch of reionization. The position of the main peak is determined by the most dominant Fourier mode stimulated at last scattering. It primarily depends on the string correlation length and the average string velocity at last scattering. Measuring the location of the main peak would provide valuable insights into fundamental physics. For example, in the case of cosmic superstrings the position of the peak of the B-mode spectrum constrains the value of the fundamental string coupling $g_s$ in string theory \cite{Avgoustidis:2011ax}.

Fluctuations sourced by strings are intrinsically non-Gaussian and hence their statistical signatures are not limited to power spectra. Several groups have made predictions for various non-Gaussian estimators that could be sensitive to cosmic strings. So far, the resulting bounds on strings are not competitive with those derived from power spectra.

The CMB bispectrum and trispectrum induced by strings on small angular scales is generally suppressed by symmetry considerations, but the trispectrum can be large \cite{Hindmarsh:2009qk,Hindmarsh:2009es,Regan:2009hv}. The trispectrum parameter $\tau_{NL}$ can be as large as $104$ for strings, hence one can anticipate strong constraints on cosmic strings as observational estimates of the trispectrum improve.

\subsubsection{21-cm}

The prospects of future observations of the $21~{\rm cm}$ cosmological background motivated investigations of their ability to constrain cosmic strings. Neutral hydrogen absorbs or emits 21 cm radiation at all times after recombination. Cosmic strings would stir the hydrogen as they move around and create wakes, leading to 21 cm brightness fluctuations. 
Under certain optimistic assumptions, future experiments with a collecting area of $10^4-10^6$ km$^2$, can in principle constrain $G\mu$ in the $10^{-10}-10^{-12}$ range \cite{Khatri:2008zw}. The same strings that create wakes would also perturb the CMB via the KSG effect, leading to potentially observable spatial correlations between the 21 cm and CMB anisotropies \cite{Berndsen:2010xc}. Also, the ionization fraction in the cosmic string wake is enhanced, leading to an excess 21 cm radiation confined to a wedge-shaped region \cite{Brandenberger:2010hn}. It remains to be seen if terrestrial and galactic foregrounds (which become very bright at low frequencies) can be overcome to use 21 cm for mapping the high redshift distribution of matter.

\subsubsection{Gravitational waves}
Oscillating loops of cosmic strings generate a stochastic gravitational wave background that is strongly non-Gaussian, and includes occasional sharp bursts due to cusps and kinks \cite{Damour:2000wa}. In the case when loops are large at formation, which is the case favored by latest simulations, pulsars currently provide the tightest bounds of $G\mu \lesssim 10^{-9}$ for ``conventional" strings, i.e. those with order unity intercommutation probability \cite{Jenet:2006sv,Blanco-Pillado:2013qja}.  (Notice that these bounds do not apply to global strings as they decay primarily through Goldstone boson radiation.)

In the case of cosmic {\it superstrings}, the probability of cusp formation is expected to be reduced due to the extra spatial dimensions, and there is some smoothing of the cusps. This can significantly damp the gravity waves emitted by cusps, and to a lesser extent by kinks, and relax pulsar timing bounds on cosmic superstrings. On the other hand, junctions on superstring loops give rise to a proliferation of sharp kinks that can amplify the gravitational wave footprint of cosmic superstrings \cite{Binetruy:2010cc}.

\subsubsection{Lensing}

The peculiar form of the metric around a cosmic strings can result in characteristic lensing patterns of distant light sources. For instance, a straight long string passing across our line of sight to a distant galaxy can produce two identical images of the same galaxy \cite{Vilenkin:1984ea}. In the more general case of loops and non-straight strings, the image patterns will be more complicated, but still have a characteristic stringy signature.

When strings bind and create junctions, as in the case of F-D superstring networks, the resulting configurations can lead to novel gravitational lensing patterns, like tripling of images when lensed by a Y-junction \cite{Shlaer:2005ry,Brandenberger:2007ae}.

The existence of cosmic strings can be strongly constrained by the next generation of gravitational lensing surveys at radio frequencies. LOFAR and SKA can give an upper bound of $G\mu < 10^{-9}$ \cite{Mack:2007ae}. Microlensing surveys are less constraining \cite{Kuijken:2007ma}.
Effects of loop clustering on microlensing \cite{Pshirkov:2009vb}, gravitational lensing due to a moving string string on pulsar timing, and quasar variability \cite{Tuntsov:2010fu} have also been considered with an aim to derive constraints. Cosmic string loops within the Milky Way can micro-lens background point sources and this offers a potentially powerful methodology for searching for cosmic strings \cite{Bloomfield:2013jka}.

Vector perturbations sourced by strings or other topological defects can generate a curl-like (or B-mode) component in the weak lensing signal which is not produced by standard density perturbations at linear order \cite{Thomas:2009bm}. Future large scale weak lensing surveys should be able to detect this signal even for string tensions an order of magnitude lower than current CMB constraints.

\subsection{Non-gravitational signatures }

In the simplest cases, such as the Abelian Higgs model, the sole impact of cosmic strings on their surroundings is through their gravity. In extended models, in which cosmic string solutions occur within a more complete particle theory, it is quite common for strings to interact via forces present in the Standard Model. However, since the precise nature of the coupling is unknown, the non-gravitational signatures of strings are more model-dependent than those discussed in earlier sections.

\subsubsection{Cosmic rays}
If strings couple to other forces, cusps and kinks can emit beams of a variety of forms of radiation which can potentially be detected on Earth as cosmic rays. For example, high energy gamma rays can be emitted from superconducting strings \cite{Vilenkin:1986zz}.

Several authors have calculated the emission of particles from strings and the possibility of detecting them as cosmic rays (for a review see \cite{Bhattacharjee:1998qc}. An important feature for certain particle-string interactions is that the flux of particles on Earth is ''inversely'' related to the string tension, at least for strings that are not too light. Thus lighter strings produce larger cosmic ray fluxes. The reason is simply that the density of string loops is greater if the strings are lighter, and the larger number of strings give a larger cosmic ray flux. Hence, if there are cosmic strings that emit cosmic rays, the constraints imply a ''lower'' bound on the string tension. Although, at very low tension, the constraint again gets weaker because then the fractional energy loss in particles is very large and this reduces the loop number density \cite{Bhattacharjee:1989vu,Long:2014lxa}. Superconducting strings can also emit high energy cosmic rays with different dependencies on the string parameters \cite{Berezinsky:2009xf}.

Another important constraint on the cosmic string scenario arises because the particles emitted by strings generally include protons and also very high energy ($\sim 10^{20}$ eV) photons \cite{Aharonian:1992qf}. Even though the nature of the ultra-high energy cosmic rays is not clear at present - they could be protons or heavy nuclei or an admixture - it is certain that they do not include a significant photon component. With particular interactions strings may be able to source the ultra-high energy cosmic rays without conflicting with the photon bounds \cite{Vachaspati:2009kq}. 

In the case of cosmic superstrings, radiation may include dilaton and other moduli. The case when the dilaton has gravitational-strength coupling to matter has been discussed in \cite{Damour:1996pv}, with constraints arising from a number of different experiments and observations. In the case of large volume and warped Type-IIB compactifications, the coupling of the moduli is stronger than gravitational-strength, and the resulting constraints in the three dimensional parameter space -- cosmic string tension, moduli mass, coupling strength -- have been analyzed in \cite{Sabancilar:2009sq}. Cosmic superstrings can also be expected to provide distinctive cosmic ray signatures via the moduli emitted from cusps.

\subsubsection{Radio bursts}
The gravitational coupling between photons and cosmic strings leads to the emission of light from strings \cite{Garriga:1989bx,JonesSmith:2009ti,Steer:2010jk}. This particular emission is generic to cosmic strings but it is suppressed by two powers of the gravitational coupling and it is unclear if it can lead to an observable signature. 

Superconducting cosmic strings --- strings that carry electric currents --- can give transient electromagnetic signatures ("radio bursts") that are most evident at radio frequencies \cite{Vachaspati:2008su}. The event rate is dominated by kink bursts in a range of parameters that are of observational interest, and can be quite high (several a day at 1 Jy flux) for a canonical set of parameters \cite{Cai:2012zd}. In the absence of events, the search for radio transients can place stringent constraints on superconducting cosmic strings, though additional recently discovered cosmological radio burst candidates are compatible with the superconducting string model \cite{Yu:2014gea}.

\subsubsection{CMB spectral distortions}
If the strings are superconducting, they emit electromagnetic radiation that produces $\mu$- and $y$-distortions of the black body spectrum of the CMB. This will allow future CMB experiments, such as PIXIE \cite{2011JCAP...07..025K}, to place tight constraints on $G\mu$ and the electric current on the string \cite{Tashiro:2012nb}.

\acknowledgments

We are grateful to Ana Achucarro, Christophe Ringeval, Paul Shellard, Gerard Verbiest and Proty Wu for permission to use figures from their publications, and to Tom Kibble and Alex Vilenkin for their comments and helpful feedback. TV acknowledges support by the DOE at ASU. LP is supported by the Natural Sciences and Engineering Research Council of Canada. DAS acknowledges the support of the excellence cluster/Labex ENIGMASS.

\bibstyle{aps}
\bibliography{biblioSch}

\end{document}